\documentclass{aa}
\usepackage{graphicx}
\usepackage{txfonts}
\begin{document}

\title{Characterization of very narrow spectral lines\\with 
  temporal intensity interferometry}


   \author{P. K. Tan
          \inst{1}
          \and
          C. Kurtsiefer
          \inst{1,2}
          }

   \institute{Centre for Quantum Technologies, 3 Science Drive 2, 117543, Singapore\\
         \and
             Department of Physics, National University of Singapore, 2 Science Drive 3, 117551, Singapore\\
             \email{pengkian@physics.org (PKT); phyck@nus.edu.sg (CK)}
             }


 
  \abstract
   {Some stellar objects exhibit very narrow spectral lines in the visible
     range additional to their blackbody radiation. Natural lasing has been
     suggested as a mechanism to explain narrow lines in
     Wolf-Rayet stars.
     However, the spectral resolution of conventional
     astronomical spectrographs is still about two orders of magnitude too low
     to test this hypothesis.}
   {We want to resolve the linewidth of narrow spectral emissions in
     starlight.
   }
   {A combination of spectral filtering with single-photon-level temporal
     correlation measurements breaks the resolution limit of
     wavelength-dispersing spectrographs by moving the linewidth measurement
   into the time domain.}
   {We demonstrate in a laboratory experiment that temporal intensity
     interferometry can determine a 20\,MHz wide linewidth of
     Doppler-broadened laser 
     light, and identify a coherent laser light contribution in a blackbody
     radiation background.}

   {} 

   \keywords{Instrumentation: Interferometers --
                Line: Identification --
                Techniques: Spectroscopic
               }

   \maketitle
%

\section{Narrow emission lines and astrophysical lasers}
Some spectral lines in the visible range emitted from stellar systems like
$\eta$ Car have a linewidth that is hard to resolve with high resolution
($10^5$) astronomical spectrographs like the Keck High Resolution Echelle
Spectrometer \citep{griest:10}.
This suggests either very low temperatures of the emission medium, or a
different mechanism like stimulated emission, which can lead to optical
emission much narrower than the participating atomic or molecular
transition \citep{schawlow:58}. 
Following first laboratory demonstrations of maser and laser radiation
\citep{maiman:60,javan:61} and  
the detection of strong interstellar microwave emission from molecular
gas clouds \citep{weaver:65},
natural non-visible lasers  from astrophysical sources were proposed to
be responsible for this emission~\citep{menzel:70, varshni:86}.

Natural stellar laser candidates in the visible range are expected to have a
spectral linewidth around 10\,MHz \citep{johansson:05,
  dravins:08, roche:12} that cannot be resolved
by conventional astronomical spectrographs.  Therefore,
alternative spectroscopical techniques like heterodyne spectroscopy
\citep{hale:00, sonnabend:05, dravins:08} or, as we 
investigate in this paper, temporal photocorrelation spectroscopy, may help to
better understand the nature of these narrow emission lines or even
verify experimentally the presence of a natural lasing mechanism in the
visible range.

\section{Intensity interferometry for time domain spectroscopy}
Intensity interferometry was used to investigate the spatial coherence properties of starlight to infer their angular diameter \citep{hbt:74}, but
first demonstration experiments were carried out on spectral lines
from a Mercury gas discharge lamp~\citep{hbt:58}. In essence, normalized
intensity correlations 
\begin{equation}\label{eq:g2def}
  g^{(2)}(\tau) = \frac{\langle I(t)I(t+\tau) \rangle}{\langle I(t) \rangle ^{2}}
\end{equation}
are recorded as a function of the time difference $\tau$ by
evaluating photodetection events from detectors observing
the same light source. 
For stationary light 
of a single polarization,
the normalized intensity correlation $g^{(2)}(\tau)$ is related to the normalized
(electrical) field correlation $g^{(1)}(\tau)$ 
\citep{mandel:95} via 
\begin{equation}\label{eq:correlationrelation}
g^{(2)}(\tau) = 1+|g^{(1)}(\tau)|^2\,.
\end{equation}
The Wiener-Khinchin theorem \citep{wiener:30, khinchin:34} links the field
correlation to the
spectral power density $S(f)$ through a Fourier transform 
$\mathcal{F}$:
\begin{equation} \label{eq:spectralamplitude}
  S(f)\propto\mathcal{F} [g^{(1)}(\tau)]\,.
\end{equation}
Therefore -- within the limits of reconstructing the phase of the complex
$g^{(1)}(\tau)$ from  $g^{(2)}(\tau)$ via
(\ref{eq:correlationrelation}) -- it is
possible to extract information about the spectral power density
$S(f)$ of the light source from a measured intensity correlation
$g^{(2)}(\tau)$. A narrow spectral distribution $S(f)$ of width $\delta f$
will result in a $g^{(2)}(\tau)$ with a characteristic time scale
$\tau_c\propto1/\delta f$.

The width $\delta f$ of narrow spectral lines can therefore be measured in the
time domain, overcoming the resolving power of wavelength-dispersive
instruments like spectrographs or narrow-band interference filters.
Note, however, that this does not allow determination of the absolute
spectral position of a line, since a frequency shift  $\Delta f$ of a narrow
distribution results in a complex oscillating term $e^{2\pi i \Delta
  f\tau}$ in $g^{(1)}(\tau)$, but leaves $g^{(2)}(\tau)$ unchanged due to
the modulus in (\ref{eq:correlationrelation}).

In stellar light sources, narrow spectral lines tend to appear
against a large background of blackbody radiation. A direct measurement of
the second order correlation function is therefore difficult, because the
signal is dominated by the blackbody contribution with a very short coherence
time on the order of $10^{-14}$\,s. Therefore, adequate
preliminary filtering has to suppress the thermal background to a level that
time domain spectroscopy can be carried out.
It is also necessary that the light exhibits some non-Poissonian
intensity fluctuations, since for light with Poissonian statistics,
e.g. coherent laser light, the intensity correlation is flat ($g^{(2)}(\tau) =
1$)  \citep{glauber:63} and has no structure that would reveal any spectral
properties.

In this work, we simulate the characteristic spectrum of natural stellar laser
candidates in the visible range by combining phase-randomized 
artificially Doppler-broadened laser light 
with spectrally wide blackbody radiation.
We then characterize the narrow spectral line by
time-resolved intensity interferometry after passing the
composite light through a diffraction grating and two etalons to suppress the
blackbody contribution.

\section{Experimental Setup}
\begin{figure}[t]
\centering
\includegraphics[width=\columnwidth]{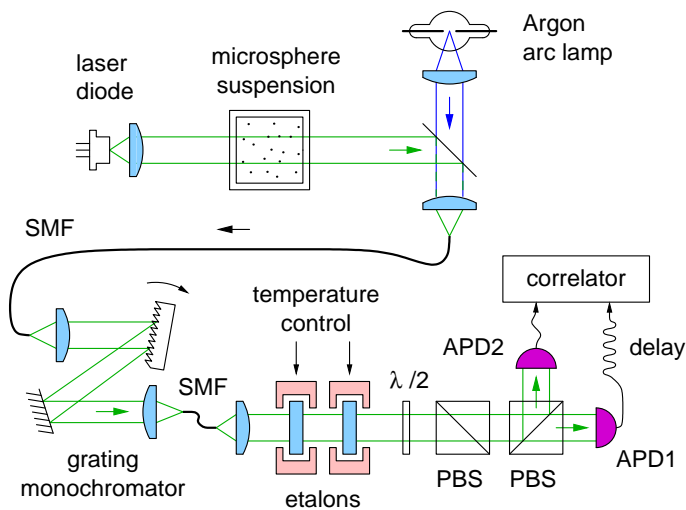}

\caption{Experimental setup. Light from a laser diode ($\lambda_L=513.8\,$nm) is
  Doppler-broadened by passing through a suspension of microspheres (0.2\,$\mu$m
  diameter), combined  with light from an Argon arc lamp on a microscope
  slide, and coupled into a single mode optical fiber (SMF). The bottom part
  shows the analysis system, consisting of a grating monochromator
  and a temperature-tuned etalon pair to select a 3.2\,GHz wide spectral window
  around 513.8\,nm from the composite light. Temporal photon pair
  correlations are recorded to identify different light
  contributions. PBS: polarizing beam splitter, $\lambda/2$: half wave plate,
  APD: single photon avalanche photodetectors.}
\label{mix}
\end{figure}

\begin{figure}[b]
\centering
\includegraphics[width=\columnwidth]{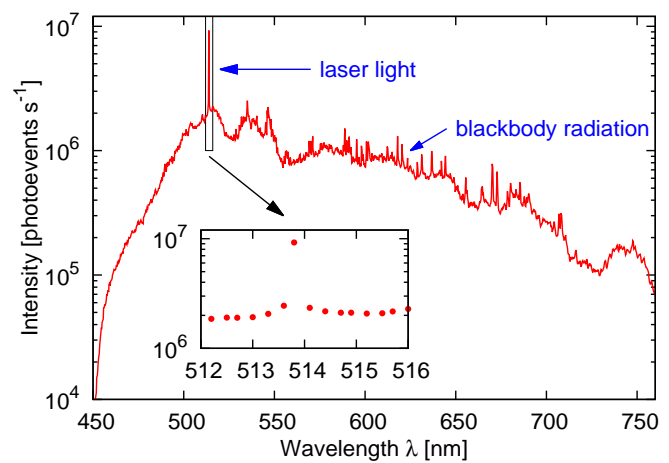}
\caption{Spectrum of the test light source in Fig.~\ref{mix} without
  Doppler broadening. The broad
  background over the whole visible range resembles blackbody radiation at an
  effective temperature $T=6000\,$K, while the inset shows the unresolved
  spectrum around the laser line. The resolution of the spectrometer is about
  0.12\,nm.
}
\label{spectrum}
\end{figure}

Our experimental setup is illustrated in Fig.~\ref{mix}.
Composite test light is prepared by combining light from a laser diode (Osram
PL520,  $P=50$\,mW) at a wavelength of $\lambda_L$=513.8\,nm with
blackbody radiation from an Argon arc lamp 
with an effective blackbody temperature of around 6000\,K 
on an uncoated microscope glass slide.
This combines approximately 4\,\% of the incident laser light with
92\,\% of the Argon arc lamp output. The resulting spectrum recorded with a
grating spectrometer of approximately 0.12\,nm resolution is shown in
Fig.~\ref{spectrum}.

The very narrowband laser light is Doppler-broadened by passing it 
through a cuvette containing a suspension of standard
mono-disperse polystyrene microspheres of 0.2\,$\mu$m diameter in water, following \citet{dravins:15}. These microspheres serve as scattering
centers undergoing Brownian motion
at room temperature. The resultant phase randomization
causes the laser light to exhibit  pseudo-thermal photon bunching behavior
\citep{martienssen:64, arecchi:65, scarl:66, scarl:68, estes:71, hard:77}.
The coherence properties of light leaving the suspension
depend on the temperature of the suspension, the viscosity (ratio of water
to microspheres), and beam focus \citep{dravins:14}; these parameters were not
fully characterized, but a combination of a beam waist of roughly 1\,mm, with beads-concentration of approximately 0.1\% solids [weight/volume] at room temperature (23 degrees Celsius) lead to Doppler-broadened light we could investigate with our technique.

The microsphere suspension with its milky appearance reduces the
intensity of the laser light by over two orders of magnitude, which is too low
to allow proper identification against the blackbody radiation background in a spectral measurement with our grating spectrometer.

To identify the laser light admixture to the blackbody radiation,
the test light is first coupled into a single mode fibre (Thorlabs 460HP).
After collimation, the
light passes through a monochromator based on a reflective diffraction
grating (1200\,lines/mm, blazed for 500\,nm). The monochromator is
calibrated
to the 546.1\,nm line from a Mercury discharge lamp where it shows
a transmission bandwidth (full width at half maximum, FWHM) of
about 0.12\,nm.

A second single mode fiber enforces spatial coherence again, before
the light passes through a pair of temperature-tuned plane-parallel solid etalons made of
fused silica (Suprasil311) with a refractive index $n$=1.4616, 
and coatings of a nominal reflectivity $R=95.2$\,\% at $\lambda_L$. 
This corresponds to an estimated finesse
${\cal{F}}_{R}=\pi\sqrt{R}/(1-R)=63.9$.
The etalons have
thicknesses of $d_1$=0.5\,mm and $d_2$=0.3\,mm, corresponding to a free
spectral range $\text{FSR}=c/(2dn)$ of 205\,GHz and 342\,GHz,
respectively. Their temperatures are stabilized 
to overlap the transmission maxima at the laser wavelength. Both etalons,
in conjunction with the diffraction grating,
suppresses most of the blackbody background \citep{pk:14}, 
transmitting only an optical bandwidth
$\delta f\approx{\text{FSR}}_1/{\cal F}_R\approx3.2$\,GHz (FWHM),
corresponding to a coherence time $\tau_{c}=1/\delta f\approx0.31$\,ns.
This filter combination has an effective spectral resolving power fo about $10^{5}$, which is comparable to current astronomical spectrographs \citep{griest:10}.

The filtered light is polarized by a first polarizing beam splitter (PBS), and
distributed by a second PBS into a pair 
of actively quenched Silicon avalanche photodetectors (APD) with a
timing jitter of about 40\,ps \citep{pk:15}. Photodetection rates are balanced
by rotating the first PBS which is preceeded by a half wave plate to maximize
the count rates.
Coincidence photoevents are recorded using
a fast digital oscilloscope. The photodetectors exhibit a dark
count rate of 50\,events/sec, predominantly from the detector thermal
noise, which is negligible in the subsequent coincidence
measurements. The coincidence histograms were normalized to obtain a
$g^{(2)}(\tau)=1$ for large $\tau$, because the oscilloscope had an unknown dead
time for histogram processing that made a direct normalization impossible.

\section{Identifying Emission Linewidth}
\label{sub:wr}
\begin{figure}
\centering
\includegraphics[width=\hsize]{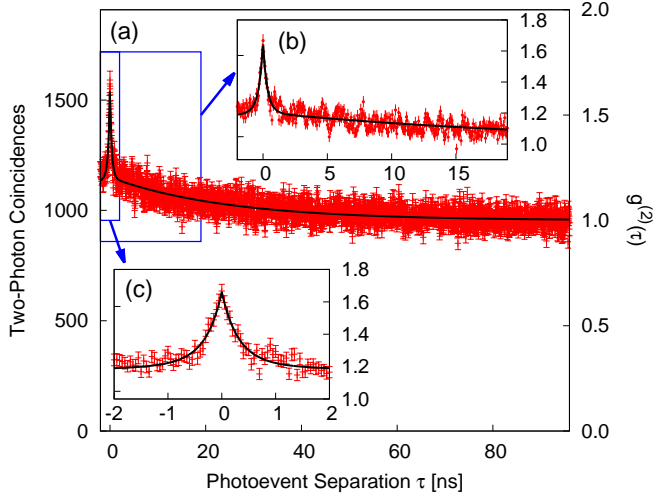}
\caption{(a) The two-photoevent coincidence histogram from filtered
  blackbody radiation with a Doppler broadened laser light contribution shows
  two exponential decays on a short and a long time scale (bin width 50\,ps).
  The solid line shows a fit of the data to model (\ref{eq:mixedmodel}),
  assuming $f_B=f_L$.
  The two zooms show (b) an oscillatory behaviour on top of the
  slow decay, and (c) a good match between the fit and the measured data for
  the filtered blackbody radiation on a short time scale. 
} 
\label{g2_arclaser}
\end{figure}

In a first experiment, we want to measure the linewidth of the laser light
that was Doppler-broadened by random scattering in the microsphere suspension
on a background of blackbody radiation. Both broadened laser light and
blackbody radiation resulted in about $2\times10^4$ photoevents per second each
behind the filter stack formed by gratings, etalons and polarization filters.

The histogram of two-photon coincidences as a function of
photodetection event separation $\tau$ is shown in Fig.~\ref{g2_arclaser}, 
with a total of $2\times10^{6}$ coincidences recorded for -2\,ns\,$<\tau<$\,96\,ns.
For time differences $|\tau|$<1\,ns, the sharp peak due to filtered blackbody
radiation is visible, while
on a longer time scale, the Doppler-broadened laser contribution due to
phase randomization in the microsphere suspension leads to photon bunching
with a slower decay constant.

A single Lorentzian frequency distribution
\begin{equation} \label{eq:lorentzianspectrum}
S(f)=\frac{\sqrt{a}}{\pi} \frac{\delta f/2}{(f-f_0)^{2}+(\delta
  f/2)^{2}}
\end{equation}
around a center frequency $f_0$ with a linewidth (FWHM) of $\delta f$
leads via (\ref{eq:correlationrelation}) and (\ref{eq:spectralamplitude}) to a normalized correlation function
\begin{equation}\label{eq:g2model}
g^{(2)}(\tau)=1+ae^{-|2\tau|/\tau_c}\quad\text{with}\quad\tau_c = 1/\delta f\,.
\end{equation}

For a mixed spectral distribution $S(f)$, the intensity correlation function
$g^{(2)}(\tau)$ can be obtained in a similar way.
If the two contributions from blackbody and
laser light are assumed to be mutually incoherent, the spectral power
densities $S_{B}(f)$ and  $S_{L}(f)$ can be added,
\begin{equation} \label{eq:summspectraldensity}
S(f)=S_{B}(f)+S_{L}(f)\,,
\end{equation}
and the resulting intensity correlation is given by
\begin{equation}\label{eq:combinedg2version1}
g^{(2)}(\tau)=1+|g^{(1)}(\tau)|^{2} = 1+  \left|\mathcal{F}^{-1}\left[S_{B}(f)\right] + \mathcal{F}^{-1}\left[S_{L}(f)\right]\right|^{2}\,,
\end{equation}
with $\mathcal{F}^{-1}$ indicating the inverse Fourier transform. Assuming now
two Lorentzian distributions $S_{B}(f)$ and  $S_{L}(f)$ according to
(\ref{eq:lorentzianspectrum}) with amplitudes $a_L$, $a_B$, coherence times
$\tau_B$, $\tau_L$, and center frequencies $f_L$, $f_b$, respectively, the Fourier transformation can easily be
carried out, leading to
\begin{eqnarray}\label{eq:mixedmodel}
g^{(2)}(\tau)&=&1+\left|a_B e^{-|\tau|/\tau_B} +  a_L
  e^{-|\tau|/\tau_L}\right|^2\\
&=&1+a_B^2e^{-|2\tau|/\tau_B}+a_L^2 e^{-|2\tau|/\tau_L}\nonumber\\
&&\quad+\,2\cos[2\pi(f_L-f_B)\tau]\,a_Ba_L\,e^{-|\tau|\left(1/\tau_B+1/\tau_L\right)}\,.\nonumber
\end{eqnarray}
For $f_L=f_B$, the oscillating term vanishes, and (\ref {eq:mixedmodel}) becomes
a sum of three exponential decays on top of $g^{(2)}=1$ that can readily explain
the correlation function in Fig.~\ref{g2_arclaser}. There, the decay for
large $\tau$ is dominated by the larger coherence time $\tau_L$. The small
peak near $\tau=0$ is a combination of two fast decays, one given by the
correlation of the blackbody contribution alone, the other one by the mixed
term with about twice the decay time for $\tau_L\gg\tau_B$.
A fit of the observed correlation function to the model (\ref{eq:mixedmodel}) 
over photoevent separations of $-2\,\text{ns}<\tau<96\,\text{ns}$ 
leads to $\tau_B=0.39\pm0.03$\,ns, $\tau_L=49.0\pm2.3$\,ns, $a_B=0.36\pm0.02$, and
$a_L=0.452\pm0.004$. However, the relatively large reduced variance
$\chi^{2}_{\text{red}}=1.26$ indicates that model (\ref{eq:mixedmodel}) is too
simple, and does e.g. not capture the oscillatory contributions in the
measured $g^{(2)}$ visible in Fig.~\ref{g2_arclaser}(b). 
The long coherence time corresponds to a linewidth of $\delta f=1/\tau_L\approx
20$\,MHz, comparable to the ones predicted for natural stellar lasers
\citep{dravins:08}. 

The described technique thus allows linewidth measurements of extremely narrow
spectral lines, limited only by the ability to record a sufficiently large
number of photons to construct a coincidence histogram. The upper bound of a
linewidth measurement with this technique is given by the time
resolution of the photodetectors and time-tagging mechanism (in our case a few
GHz). However, the phase uncertainty of $g^{(1)}(\tau)$, if inferred from
$g^{(2)}(\tau)$ in (\ref{eq:correlationrelation}), requires further assumptions
for a direct reconstruction of a spectrum via (\ref{eq:spectralamplitude}).

\section{Identifying Coherent Light}
\label{sub:seti}
\begin{figure}
\centering
\includegraphics[width=\hsize]{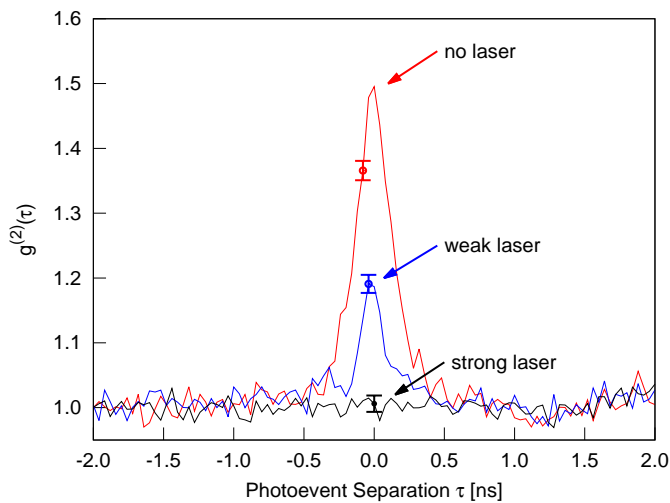}
\caption{Temporal photodetection correlations for different ratios of
  coherent laser and filtered blackbody radiation:
  all measurements have a blackbody contribution
  of approximately $3\times10^{4}$ photoevents/sec. For the ``strong
  laser'' trace, the laser contributed about  $6\times10^{6}$
  photoevents/sec, for the ``weak laser'' trace about 
  $3\times10^{4}$ photoevents/sec. For reference, the photodetection
  correlations of filtered blackbody radiation without any laser light is also
  shown. Each measurement accumulated $10^{6}$ coincidence
  photoevents with -3.1\,ns$<\tau<$3.3\,ns into 40\,ps wide bins to allow for
  direct comparison of the resulting histograms. The error bars reflect
  Poissonian counting statistics and  are representative for all time
  differences. Fitting the ``no laser'' trace to model (\ref{eq:g2model})
  leads to a coherence time $\tau_c$=0.31$\pm$0.01\,ns, and to
  $\tau_c$=0.26$\pm$0.03\,ns for the trace with a weak laser.} 
\label{g2_etalon}
\end{figure}

In a second experiment, we try to identify the presence of coherent laser
emission by a quantitative evaluation of the photobunching signature
$g^{(2)}(\tau=0)$. For this, we remove the microsphere suspension, and record
the temporal correlation measurement for different admixture levels of
attenuated laser radiation to a blackbody radiation background of about
$3\times10^{4}$ photoevents/sec after the filter stack. Assuming a Lorentzian
spectral distribution (\ref{eq:lorentzianspectrum}), the fit of the observed
second order correlation leads to a a coherence time
$\tau_c$=0.31$\pm$0.01\,ns, in agreement with $\tau_B$ obtained
from the fit in the first experiment.

The results are shown in Fig.~\ref{g2_etalon}. Without any laser light
contribution, a detector-limited blackbody temporal bunching signature of
approximately $g^{(2)}(0)=1.5$ is observed, compatible with the transmission
bandwidth around 3.2GHz of the etalon stack at $\lambda_L$ central wavelength
and the timing jitter of the avalanche photodetectors \citep{pk:15}.

For a weak laser contribution ($\approx10^{4}$ photoevents/sec) on top of a
blackbody background, the temporal photon bunching signal is reduced to
$g^{(2)}(0)\approx1.2$, indicating a sub-thermal photon bunching
signature. This means that even the presence of small contributions of
coherent light is revealed by the reduction of the thermal photon
bunching signature expected from the filtered blackbody component.

For the third trace in Fig.~\ref{g2_etalon}, the  laser light
contribution is over two orders of magnitude stronger than the filtered
blackbody contribution, corresponding to the power ratio used to obtain
the spectrum in Fig.~\ref{spectrum}.  The timing correlation
appears constant within the statistical uncertainty, without an observable
temporal photon bunching signature from the blackbody contribution.

The last trace resembles a typical photodetection correlation observed
photodetectors exposed to wideband radiation, like in the traditional
experiments of \cite{hbt:58}, but with a significant
difference: since the optical bandwidth of the detected radiation is narrower
than the inverse detector timing uncertainty, the {\em reduction} of a
photobunching signal can be interpreted as a signature of a
light source with sub-thermal statistics, e.g. due to contributions of
coherent light from a lasing mechanism.

\section{Summary}
Time-resolved second order correlation spectroscopy was used to identify the
presence of very narrow-band light on a thermal background. The
linewidth of pseudo-thermal light could be determined that was generated by
phase-randomization in a multiple scattering process, similar to light from an
ensemble of emitters without a fixed phase relationship, like a gas cloud
excited by a nearby star.
Temporal intensity interferometry offers a spectral resolution of at least a
few 10\,MHz for emission lines, exceeding by far that of contemporary
astrophysical spectrographs \citep{griest:10}.

Also, an identification of sub-thermal photon statistics can be carried out
with the presented technique indicating a possible
optical lasing mechanism, and therefore help to better understand
the very narrow spectral features of stellar light sources even
in presence of a strong blackbody radiation background.


\begin{acknowledgements}
We acknowledge the support of this work by the National Research Foundation
and the Ministry of Education in Singapore, partly through the Academic
Research Fund MOE2012-T3-1-009.
\end{acknowledgements}


\bibliographystyle{aa}

\end{document}